# Variability of F2-layer peak characteristics at low latitude in Argentina for high and low solar activity and comparison with the IRI-2016 model

## González, G. de L.[1*] and López, J.[2]

*1. Assistant Professor, Department of Physics, Faculty of of Ingineering, Saint Thomas Aquina University of the North, Tucumán, Argentina*

*2. Researcher, CIASUR - National Technological University, Tucumán, Argentina*

## Abstract

This work presents the study of the variability of foF2 and hmF2 at a low latitude station in South America (Tucumán, 26.9°S, 294.6°E; magnetic latitude 15.5°S, Argentina). Ground based ionosonde measurements obtained during different seasonal and solar activity conditions (a year of low solar activity, 2009 and one of high solar activity, 2016) are considered in order to compare the ionospheric behavior. The parameters used to analyze the variability are the median, upper and lower quartiles. In addition, the foF2 values are compared with those estimated by the International Reference Ionosphere (IRI) - 2016 model. It is found that: a) A clear dependence on solar activity is observed in foF2 and hmF2, both increase with increase in solar activity. b) the variability of foF2 is higher at low solar activity, this behavior is not observed in hmF2 that present similar variability during both periods. c) the variability of foF2 is larger at night than during the day, this behavior is more pronounced during the high solar activity period. d) The variability of foF2 is higher than that of hmF2. e) Significant planetary wave spectral peaks at about 2 and 5 days are observed at high and low solar activity. f) In general, IRI overestimates foF2 during daytime, and underestimates it at post-sunset period, a better agreement is shown during nighttime.

## 1. Introduction

The ionosphere is the ionized region of the atmosphere that extends from 60 km up to 1000 km. It is divided into different regions or layers: D, E, F1 and F2, according to the neutral composition and the source of ionization at different heights (Kelley, 2009). The ionosphere is a dispersive medium that causes a time delay or phase advance in radio signals of ground-space communications, Global Navigation Satellite System (GNSS), satellite altimeters, and space-based radars (Sales, 1992; Davies, 2008). The ionospheric parameters such as electron density, ion composition, temperature, etc. vary with the solar-cycle. This complex variability is not completely understood. The F2 layer (250 km above sea level) has the highest electron density in the ionosphere; therefore, the variation of the F2-region electron density distribution greatly affects the propagation of the radio waves. The maximum electron density of the ionosphere, NmF2, occurs at the peak height hmF2, and is related to the F2 critical frequency, foF2, as NmF2 = 1.24 x $10^{10}$foF2$^2$. NmF2 is in electrons per cubic meter and foF2 is in megahertz. The variations in foF2 indicate the events occurring there.

A good description of the variability of ionospheric magnitudes, such as the F2-layer peak characteristics foF2 and hmF2, is perhaps the biggest challenge for ionospheric forecasters and is needed to improve the performance of the ionospheric models. The forecasting methods also depend on the understanding of the physics of the magnetosphere-ionosphere-thermosphere coupling mechanisms and on successful solar and magnetospheric predictions. The aim of the forecasters is to predict the day to day variability of the ionosphere so advances in ionospheric research are fundamental, especially at the F layer, which is the principal reflecting layer for long distance HF communications and navigation (Altinay et al., 1997; Kumluca et al., 1999; Meza et al., 2000; Strangeways et al., 2009; Shim et al., 2012). Many techniques have been developed for ionospheric forecasting: empirical approaches based on statistical models such as the autocorrelation method (ACM) or the Geomagnetically Correlated Autoregression Model (GCAM) (Mikhailov et al., 2007); artificial neural network (ANN) methods that use time series prediction capabilities of artificial intelligence techniques for ionospheric prediction, for example the neural network based autoregressive model with additional inputs (X) (NNARX)(Altinay et al., 1997; Kumluca et al., 1999; Wintoft and Cander, 2000; Oronsaye et al., 2014); Forecasting maps such as the Short-Term Ionospheric Forecasting (STIF) an operational tool based on continuous monitoring of the ionosphere and developed at Rutherford Appleton Laboratory, Chilton, UK, that produces forecast maps of ionospheric parameters for the European region.

*Corresponding author:                                                                      gildadelourdes@gmail.com

The waves in the neutral atmosphere, their interactions and modulations also affect the ionosphere and the ionospheric predictions. Therefore, it is useful to know when it is necessary to take the wave-type oscillations into account. These waves are: planetary waves with periods of about 2–30 days, tidal waves with periods of 24 and 12 hours for solar tides and 14.75 days for the lunar semidiurnal tide, gravity waves with periods between several minutes and a few hours and infrasonic waves with periods from about 1 s to a few minutes (Whitehead, 1971; Laštovička, 2006; Alam Kherani et al., 2009; Eccles et al., 2011; Abdu, 2016). The waves are an important component of low-latitude ionosphere variability and their impact can be studied using ionospheric parameters such as foF2, h´F (minimum virtual height of the F trace), hmF2 or NmF2 (Ogawa et al., 2006; Chum et al., 2014; de Abreu et al., 2014).

In this work, the International Reference Ionosphere (IRI) model is used. It is a data-based empirical model and is an international standard for the parameters in the earth´s ionosphere. This model is based on worldwide available data from various sources like ionosondes, incoherent scatter radars, rockets and satellites. It is maintained and revised by an international working group established by the Committee on Space Research (COSPAR) and the International Union of Radio Science (URSI) (Bilitza et al., 2011). Its development started in 1978, since then many improvements have been made, its latest version is IRI-2016 (Bilitza et al., 2017).

It is also important to consider the location (latitude and longitude) of the ionosonde used for the analysis since some models do not have good performance in all latitudes. Latitude is the angular distance from the solar equator, measured north or south along the meridian and longitude is the angular distance from a standard meridian (0 degrees heliographic longitude), measured from east to west (0 to 360 degrees) along the Sun's equator. In aeronomy, it is habitual to use the geomagnetic latitude; this coordinate has the same relation to the geomagnetic dipole equator as geographic latitude does to the geographic equator. According to this coordinate, the ionospheric behavior is divided into three main regions: low latitudes where geomagnetic latitude is between 0° and 20°on each side of the magnetic equator, mid latitudes where geomagnetic latitude is between 20° and 60° on each side of the magnetic equator and high latitude where geomagnetic latitude is between 60° and 90° on each side of the magnetic equator (Zolesi and Cander, 2014). The IRI model offers accurate modeled ionospheric parameters at mid-latitudes, but comparison with observational data at low latitudes has brought out some discrepancies.

In the present work, variations in foF2 and hmF2 data from an ionosonde station at Tucumán-Argentina (26.9 ° S, 294.6 ° E, lat Geomagnetic 15.5°S) are examined during two different solar-activity periods: low solar activity and high solar activity, by using the median and quartiles. Furthermore, atmospheric waves effects in the ionosphere are identified in the ionosonde data. Finally, foF2 values are compared with the corresponding output given by the IRI.

## 2. Data and Method of analysis

Several indexes have been used to study the variability of foF2. The most common methods adopted are the mean ($\mu$) and the standard deviation ($\sigma$) in case of a normal distribution; however, if the distribution is not normal the upper quartile ($Q_{up}$) and the lower quartile ($Q_{lo}$) are used to specify the dispersion from the central value (median) (Spiegel, 1976). The standard deviation is a good measure to describe the variability of the ionosphere, but it is very difficult to interpret in terms of probability since you cannot be sure that the distribution of the data is a Gaussian (Bilitza et al., 2004). Using the quartiles, the probability that the foF2 value falling between lower quartile and median is 25% and the probability that the foF2 value falling between median and upper quartile is 25%. In addition, the median and quartiles are less affected by magnetic storms (Ezquer et al., 2004).

Since the data behaves as a non-normal distribution, the following variability parameters are used for the analysis:

- The monthly medians m, the upper quartile $Q_{up}$ and lower quartile $Q_{lo}$
- the interquartile range $Qr = Q_{up} - Q_{lo}$
- and the variability indexes $C_{up} = Q_{up} / m$, $C_{lo} = Q_{lo} / m$ and $Cr = C_{up} - C_{lo}$.

For example, $C_{up} = 1.20$ means that $Q_{up}$ is 20% greater than median and, $C_{lo} = 0.80$ means that $Q_{lo}$ is 20% lower than median. The variability is low when $C_{up}$ and $C_{lo}$ are close to 1 and is high when the indexes are far from 1.

The data used in this study are the ionospheric parameters foF2 and hmF2 at Tucumán, obtained with the "Advanced Ionospheric Sounder" (AIS). This ionosonde sweeps between 1 and 25 MHz in 30 seconds every 5 minutes. The authors do not have information about the availability of this type of instrument in Iran. However, the study of ionospheric variability can be done using GNSS satellites that have worldwide coverage.

The median is calculated per hour for each month analyzed, only those hours for which there are more than 10 measurements in the month are considered. The data are arranged in two data sets to analyze the ionospheric variations according to the solar activity. The low solar activity (LSA) data set using the 2009 ($Rz_{medium} = 4.8$) data and the high solar activity (HSA) data set using 2016 ($Rz_{medium} = 39.8$) data. Figure 1 shows the temporal variability of monthly averages of Rz for 2009 and 2016. To examine the seasonal effects, all data is divided into three groups: winter, autumn and spring using three months of data for each season, [i.e. autumn (March, April and May), spring (September, October and December) and winter (June, July and August)]. Not enough data is available for January, February and November, so these months are not considered in the study.

## 3. Results
### 3-1. Analysis of foF2 and hmF2
Figures 2, 3 and 4 show the temporal variation of foF2 at Tucumán during months of autumn, spring and winter of 2009 and 2016. It is observed that in all cases the highest values of foF2 correspond to 2016, that is, high solar activity. The monthly variation trends are, in general, the same under both LSA and HSA: foF2 reaches the lowest level, just before dawn (between 05LT and 07LT), increases rapidly after sunrise due to the photoionization, it is amplified during the day and then decreases in the evening.

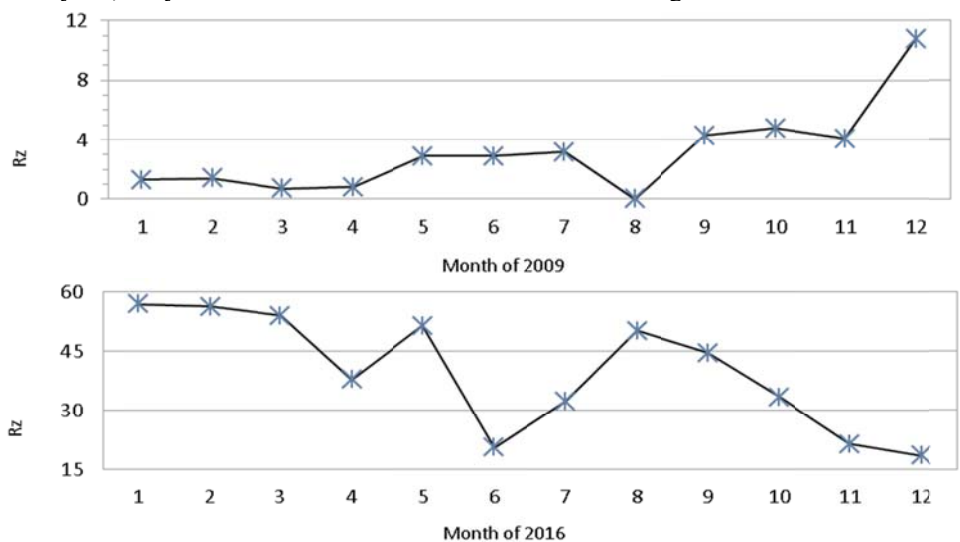

**Figure 1.** Average monthly values of Rz for 2009 (a, LSA) and 2016 (b, HSA). Source: WDC-SILSO, Royal Observatory of Belgium, Brussels.

Regarding the maximum values, two cases are distinguished; months with two peaks and months with one peak. April presents maximums at 13LT and at 17LT for low and high solar activity periods, May shows peaks at 12LT and at 17LT under HSA and at 13LT and 17LT under LSA, July presents two peaks at 12 LT and at 16 LT for HSA and one for LSA at 15 LT. The first peak (12 – 13 LT) is produced because the sun is at maximum incidence, while the second one (16 – 17 LT) appears at the time when the fountain effect has the maximum accumulation of plasma on the south crest of the equatorial anomaly. In the other months a single peak is observed, March presents one at 18LT under HSA and at 19LT under LSA; June has a peak at 16LT during low and high solar activity periods; August has a peak at 16LT under HSA and at 15LT under LSA; September has a maximum at 17LT during both periods; October presents a maximum at 18LT under HSA and at 19LT under LSA, finally, for December there is a peak at 17LT under HSA and at 18LT under LSA, in addition, foF2 values are very similar during both periods. During winter, the peak observed between 16 and 19 LT is less pronounced; this is probable because in this period the vertical drift at equatorial latitudes is smaller than during other seasons especially for LSA, consequently the fountain effect is reduced.

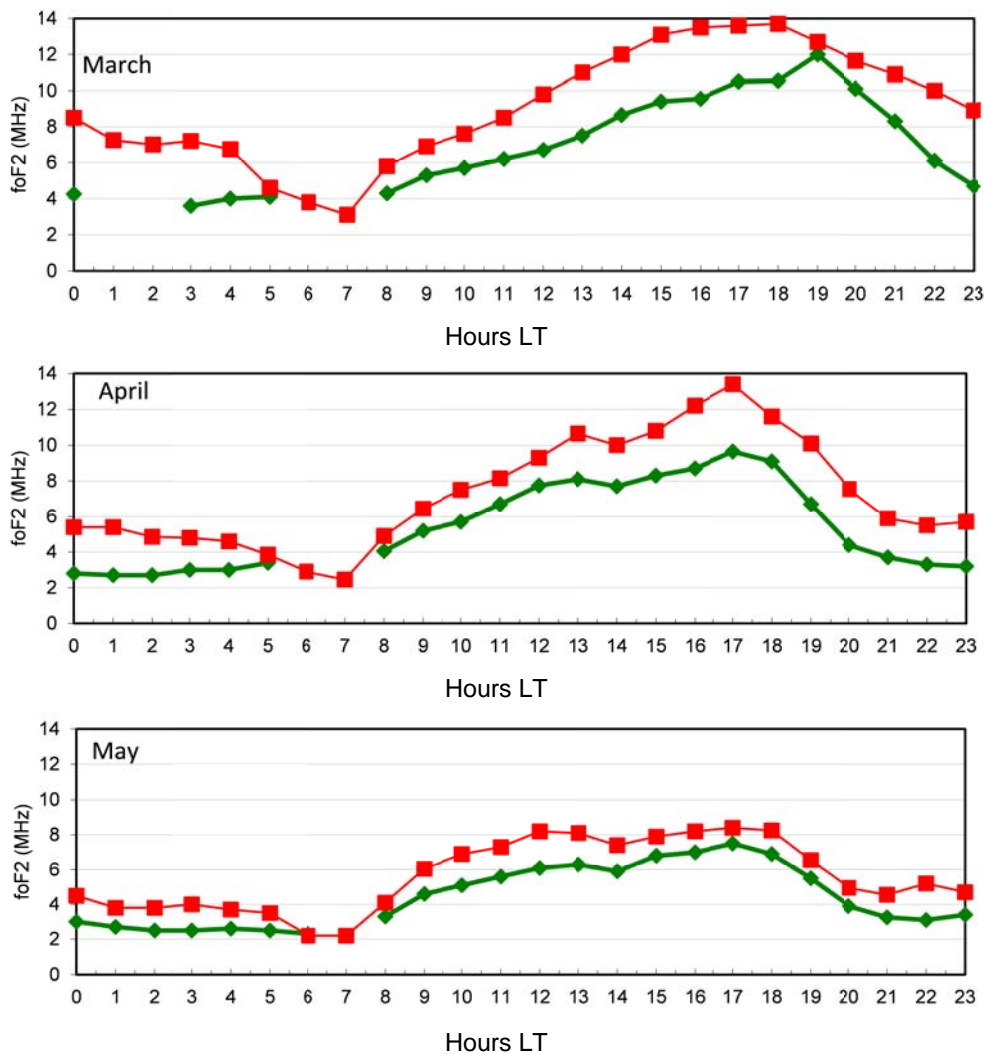

**Figure 2.** Average foF2 observed in Tucumán at autumn, 2016 (red) and 2009 (green). The X axis represents the hours in local time (LT).

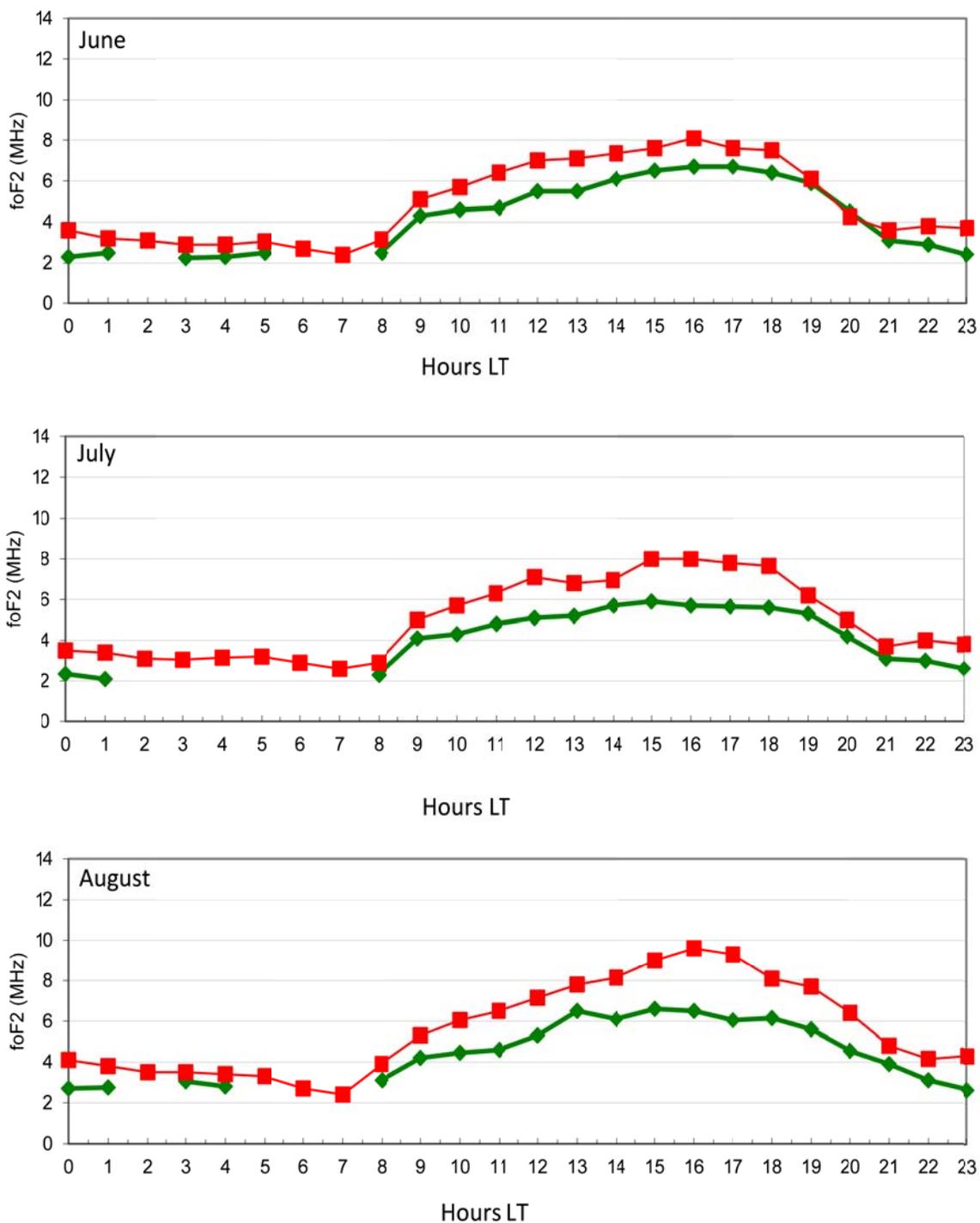

**Figure 3.** Average foF2 observed in Tucumán in winter, 2016 (red) and 2009 (green).

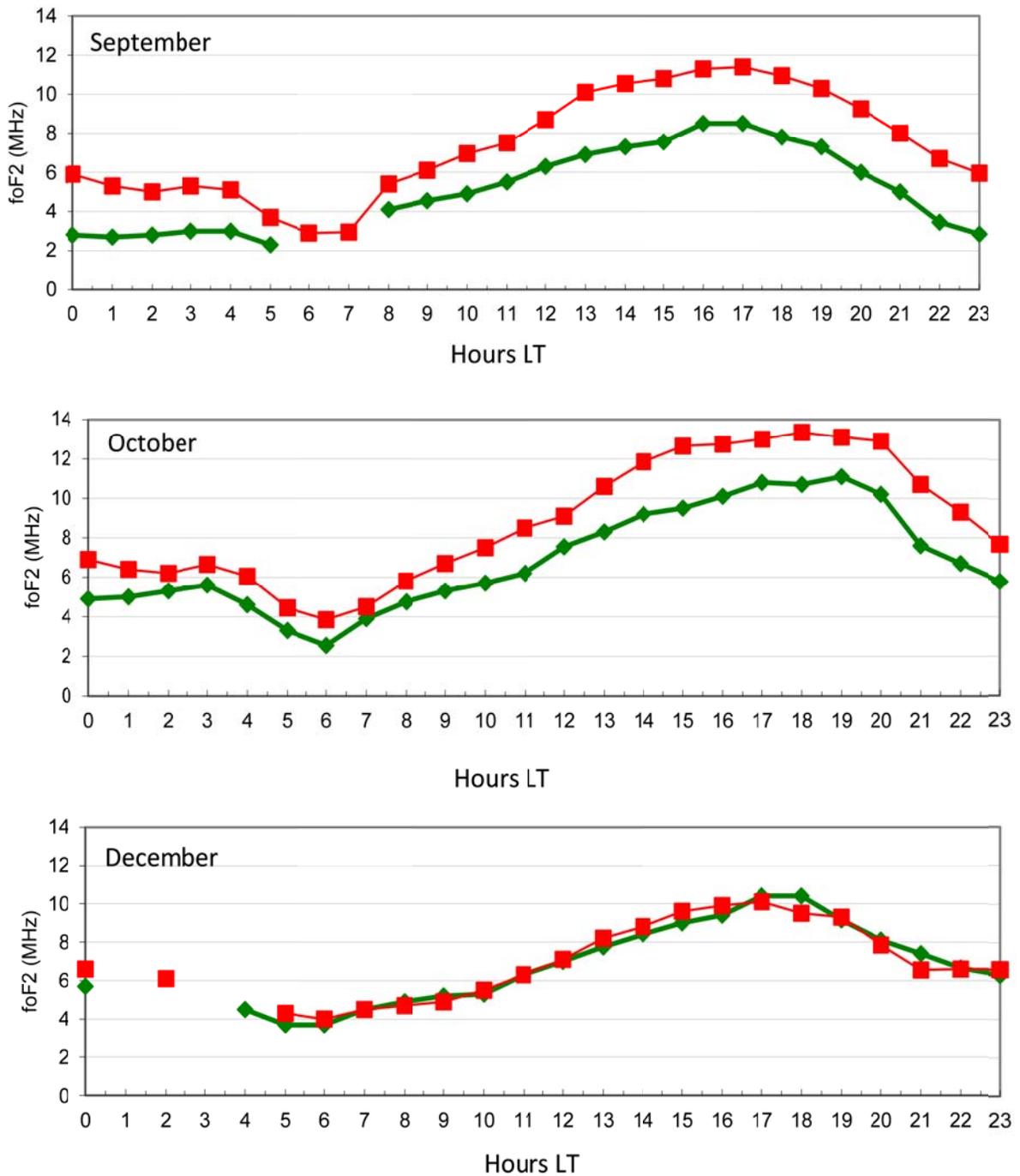

**Figure 4.** Average foF2 observed in Tucumán in spring, 2016 (red) and 2009 (green).

In general, winter shows lower foF2 values than equinox during the LSA and the HSA periods, this phenomenon is called semiannual anomaly. It reflects the seasonal nature of winter, when less solar radiation reaches the Earth, and there are fewer daylight hours. It could also be caused by changes in the neutral gas composition of the upper atmosphere, the increase in the molecular nitrogen density and the decrease in the atomic oxygen density may contribute to reduce the ionization density at peak F2-region height. Moreover, hmF2 vs. local time is shown in Figures 5, 6 and 7. In general, the highest hmF2 values occur during HSA specially in spring and the lowest in winter during LSA. The biggest difference between HSA and LSA happens in March however the general trend is similar. For March, September, October and December there are two peaks around 4 and 16 LT for both periods. Whereas for the other months, especially in winter, the first peak is less pronounced and the second one almost disappears. There is a third peak at 8 – 10 LT that is more intense at equinox during HSA and is not present in May LSA nor in June. It is also observed that during winter the two curves (HSA and LSA) are close and they overlap in some intervals.

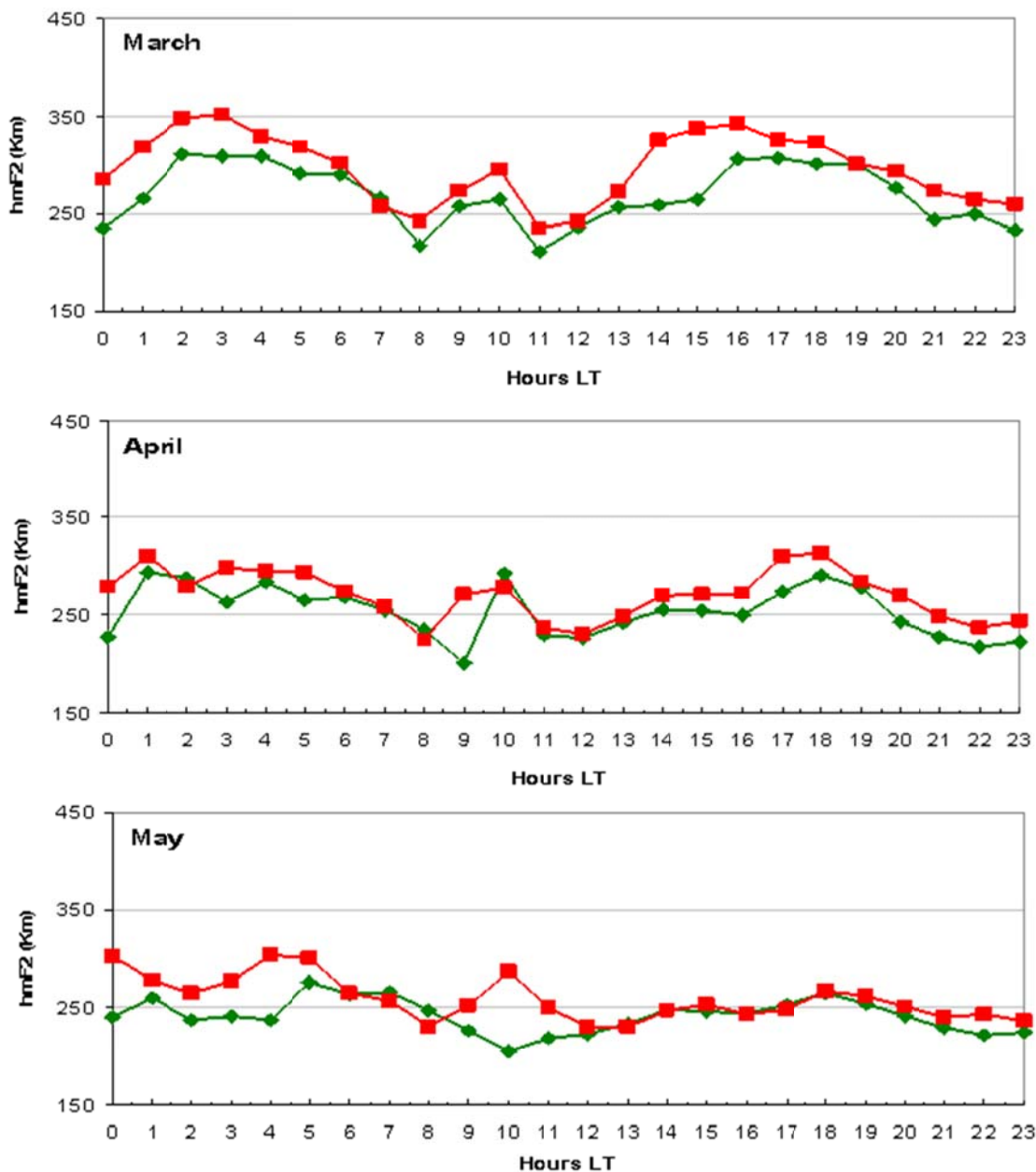

**Figure 5.** Average hmF2 observed in Tucumán in autumn, 2016 (red) and 2009 (green).

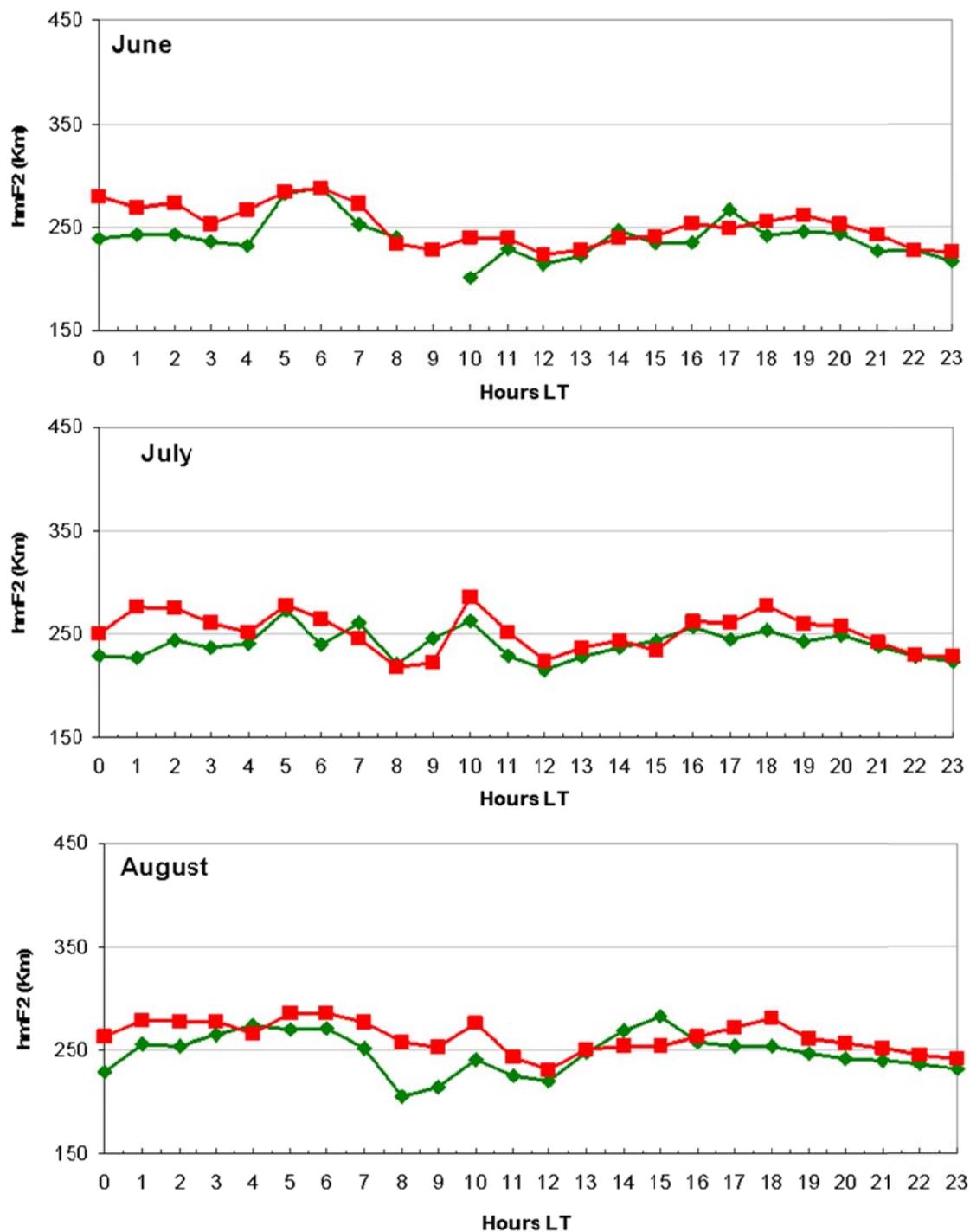

**Figure 6.** Average hmF2 observed in Tucumán in winter, 2016 (red) and 2009 (green).

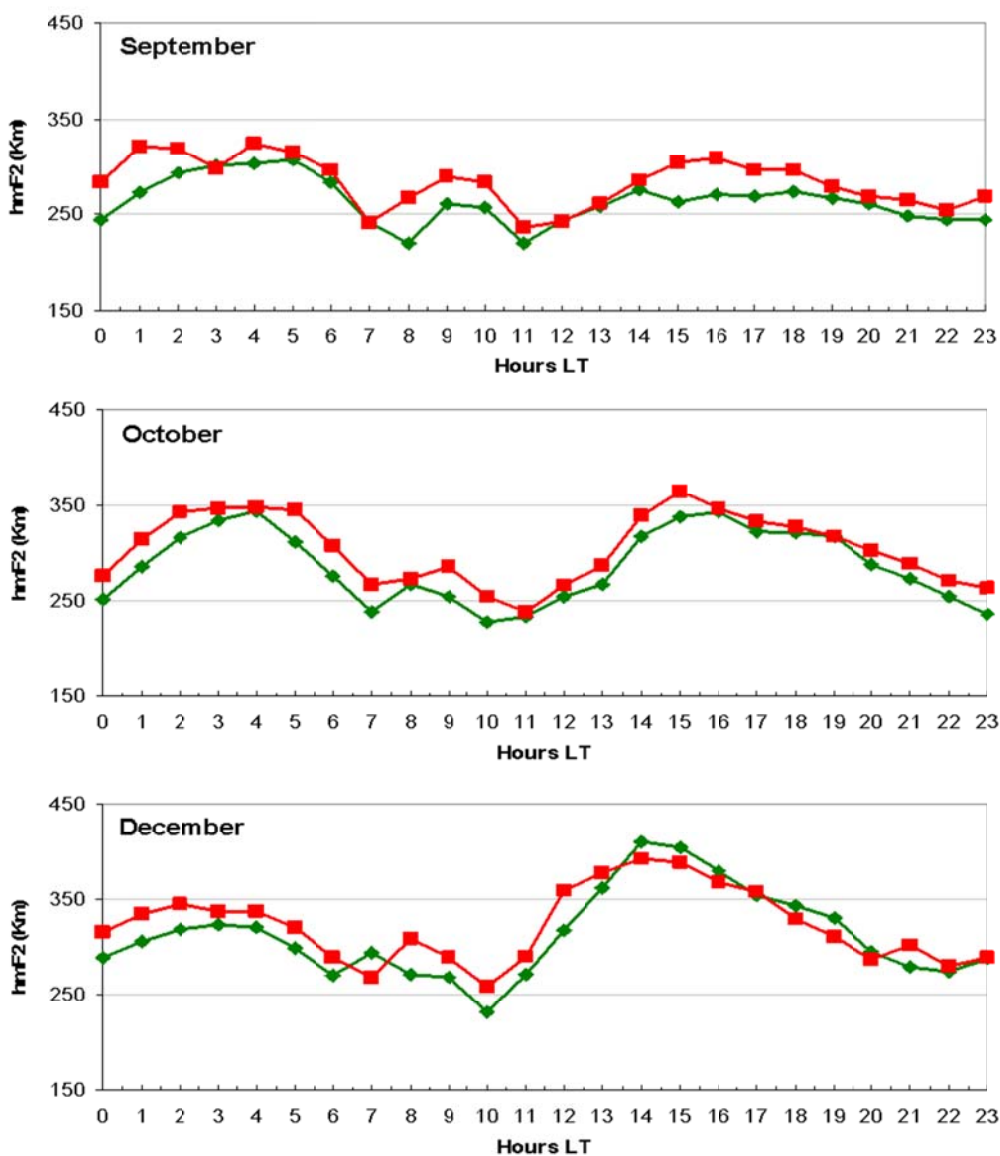

**Figure 7.** Average hmF2 observed in Tucumán in spring, 2016 (red) and 2009 (green).

For the two ionospheric parameters analyzed here, foF2 and hmF2, the values obtained during HSA are higher than those measured during LSA although the general behavior is similar during both periods. Moreover, the two magnitudes show the lowest values during winter and the highest values during equinox.

The second analysis is carried out through the interquartile difference Cr ($C_{up}-C_{lo}$) to identify higher and lower variability conditions. With a model like this, it is possible to predict that the foF2 or hmF2 value is between the lower quartile and the upper quartile with 50% probability, that is, the difference between the upper quartile and the lower quartile covers 50% of the data. The obvious disadvantage is that the quartiles ignore the other 50% of the data.

Figures 8, 9 and 10 show the temporal variation of $C_{up}$, $C_{lo}$ and Cr for foF2 for each month during LSH and HSA periods and Figures 11, 12 and 13 show the variability index for hmF2. For July and August, there is not enough data between 0LT and 9LT for 2009 and 2016 so the variability in that period cannot be analyzed. It was observed that, generally, the lowest variability occurs between 8LT and 11LT and the greatest variability between 19LT and 07LT, thus, the variability is higher during the night than during the day. In April, there are three variability peaks under high solar activity, at 14LT, 18LT and at 20LT. In May, a peak is observed for LSA period at 17LT. In July, the peaks are observed at 14LT and at 20LT under low solar activity and at 2LT, at 6LT and at 12LT under high solar activity. During July, the variability was not higher at night than day as was observed for other months. In August, there is a peak at 20LT under high solar activity and at 17LT under low solar activity. In October, a peak is observed at 4LT under low solar activity and at 5LT under high solar activity. At night during LSA period, October presents the highest Cr values of all analyzed months while the lowest Cr values happen in December. It is also observed that the variability is generally lower during winter and is higher during the LSA period, especially for September and October.

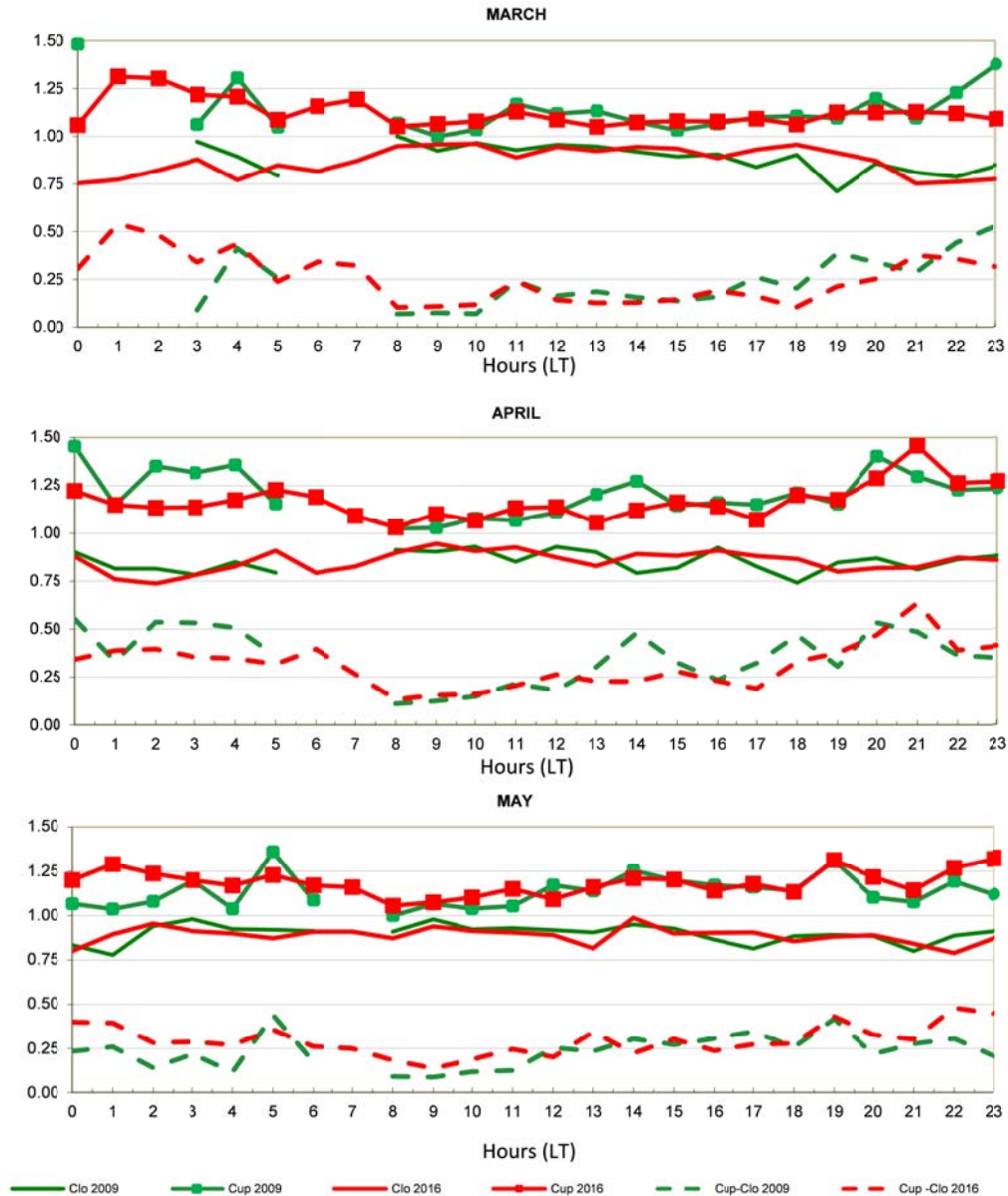

**Figure 8.** Temporal variability of upper and lower quartiles and interquartile range for foF2 for one year of high solar activity (2016) and one of low solar activity (2009).

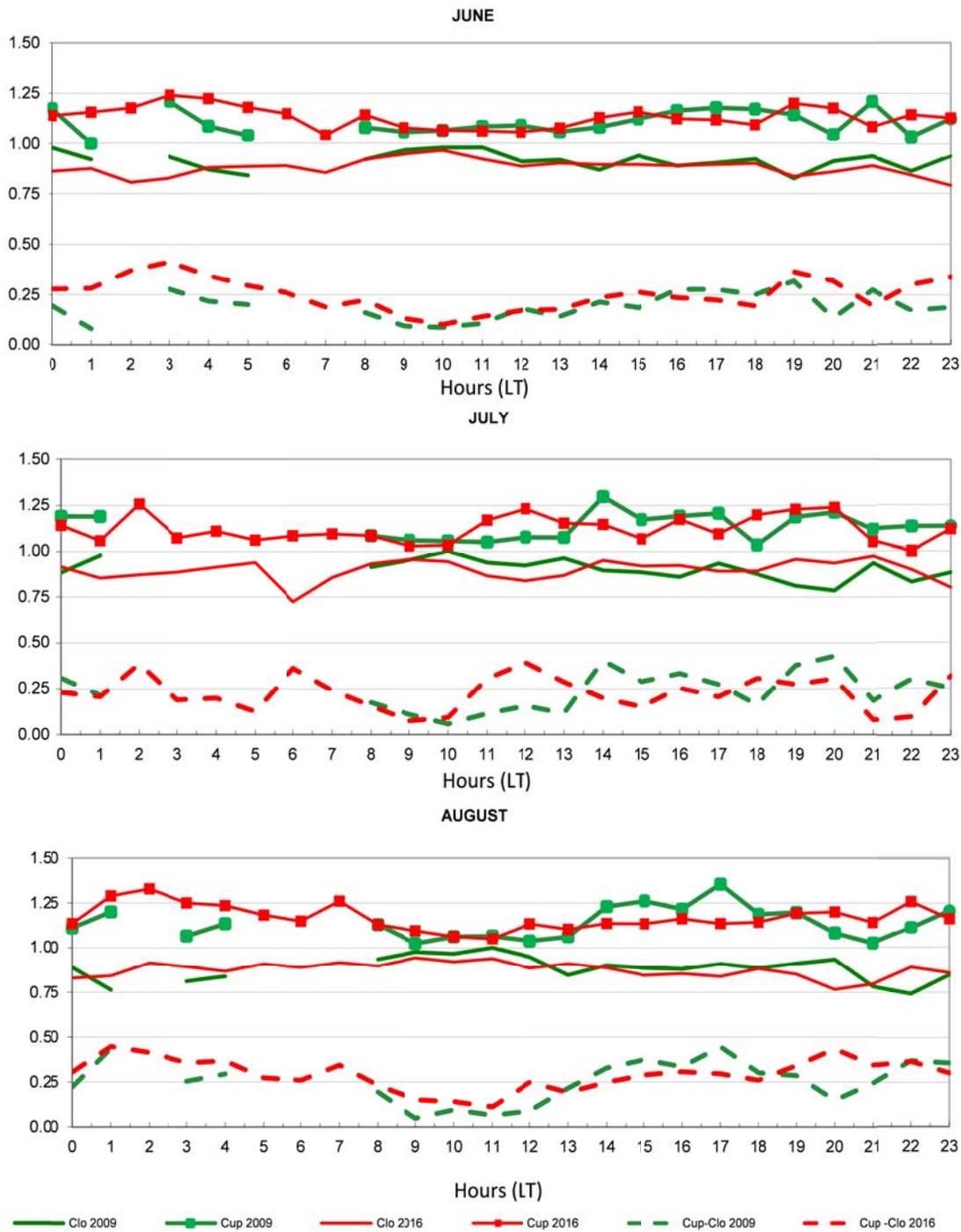

**Figure 9.** Temporal variability of upper and lower quartiles and interquartile range for foF2 for one year of high solar activity (2016) and one of low solar activity (2009).

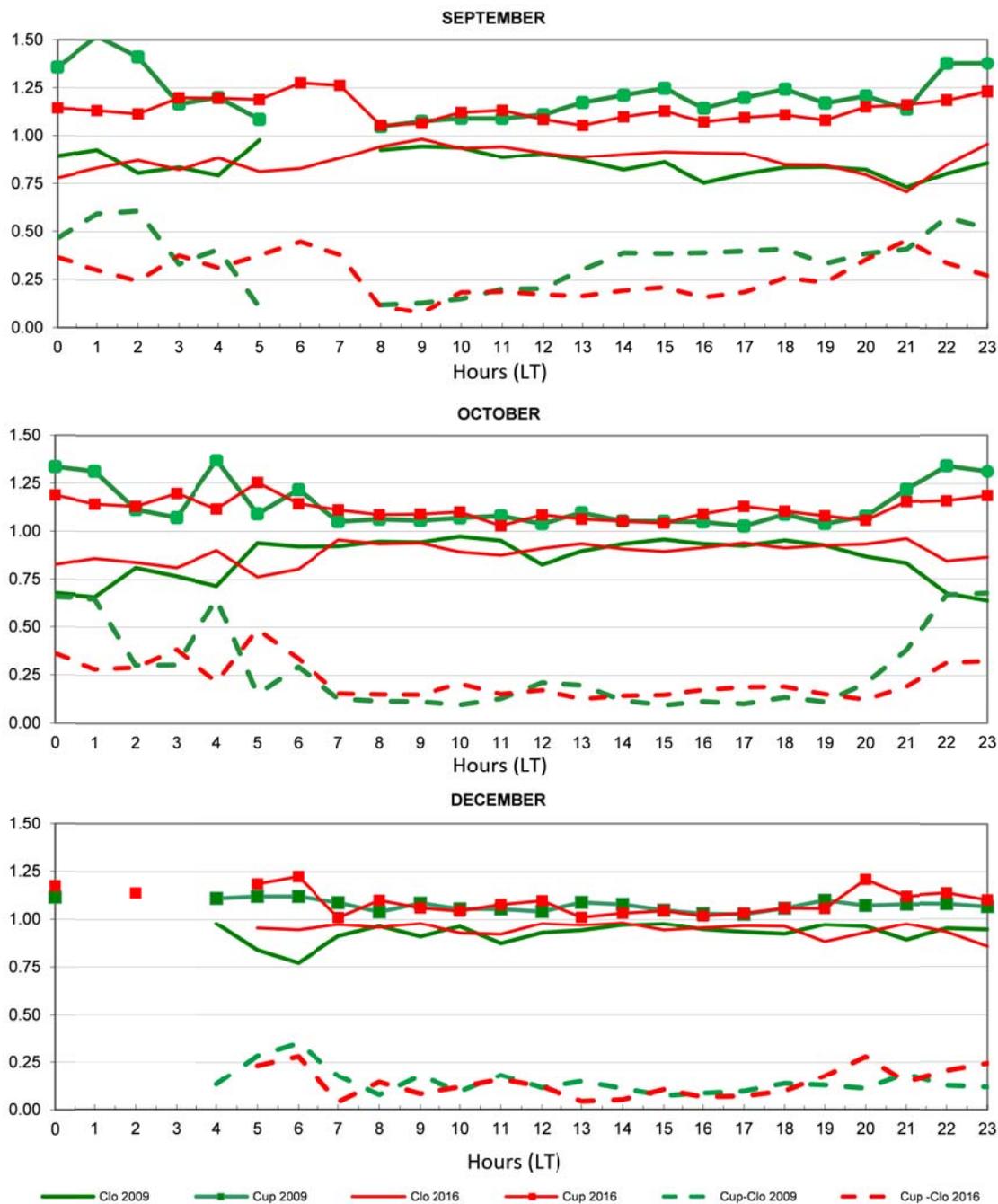

**Figure 10.** Temporal variability of upper and lower quartiles and interquartile range for foF2 for one year of high solar
activity (2016) and one of low solar activity (2009).

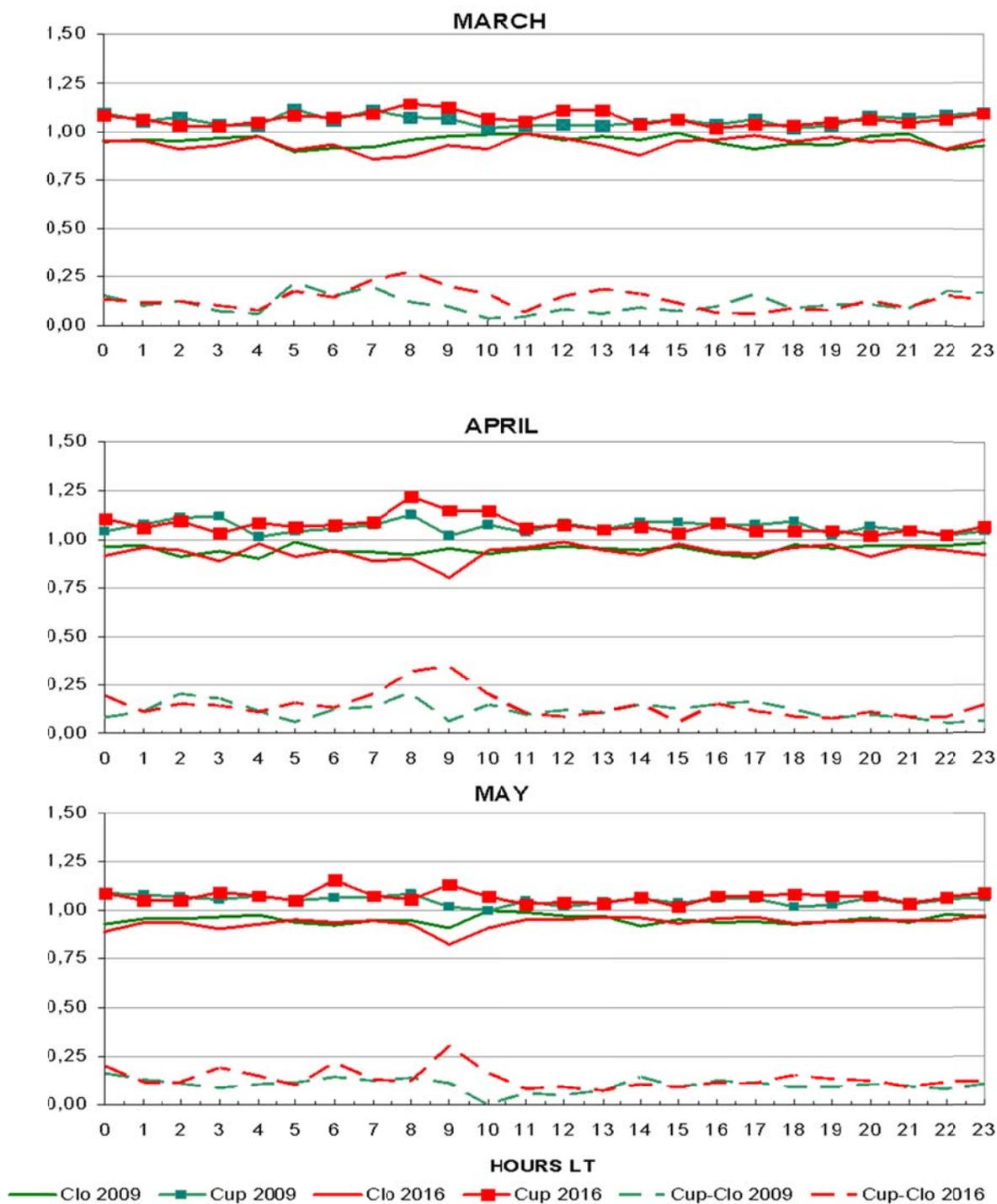

**Figure 11.** Temporal variability of upper and lower quartiles and interquartile range for hmF2 for one year of high solar activity (2016) and one of low solar activity (2009).

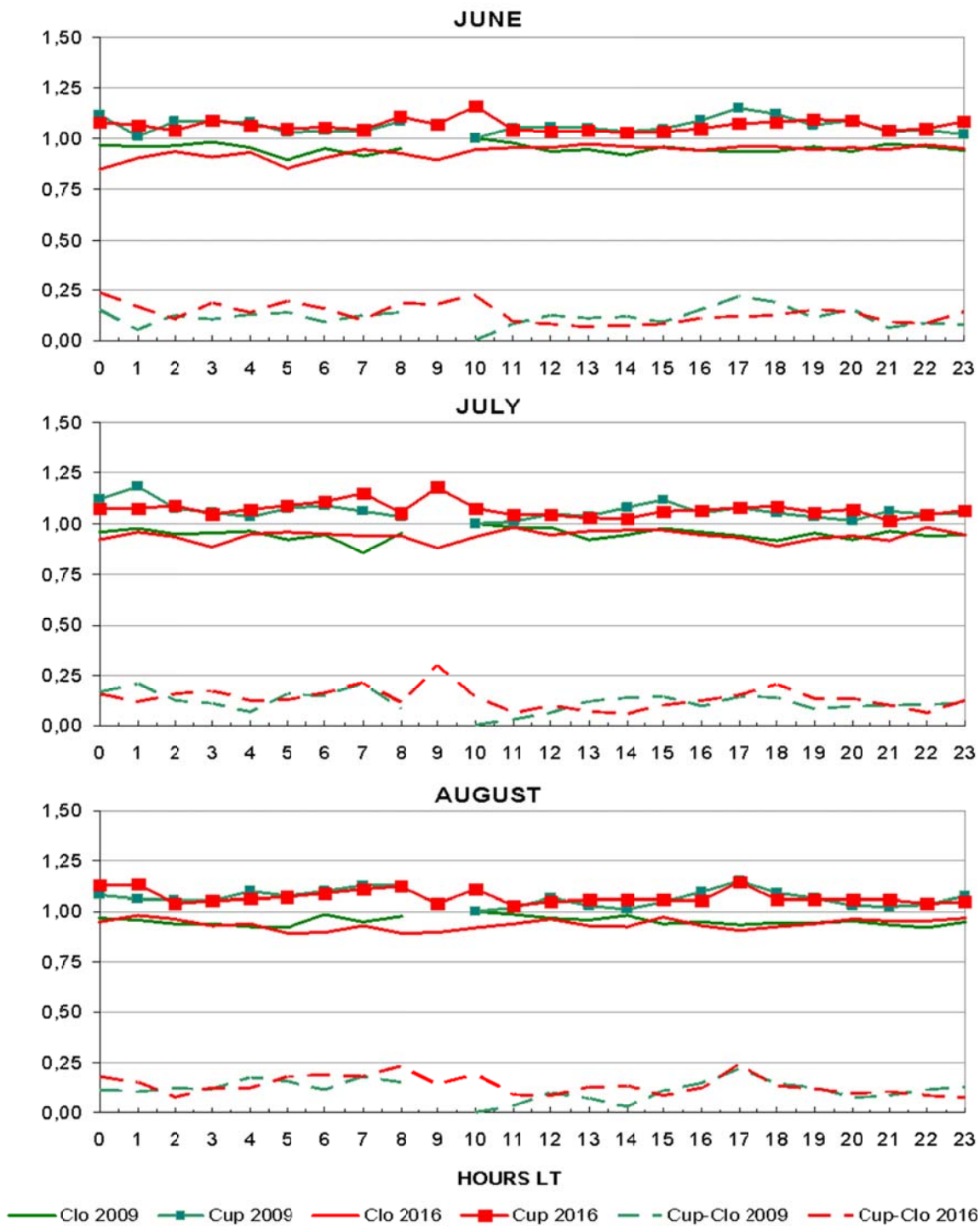

**Figure 12.** Temporal variability of upper and lower quartiles and interquartile range for hmF2 for one year of high solar activity (2016) and one of low solar activity (2009).

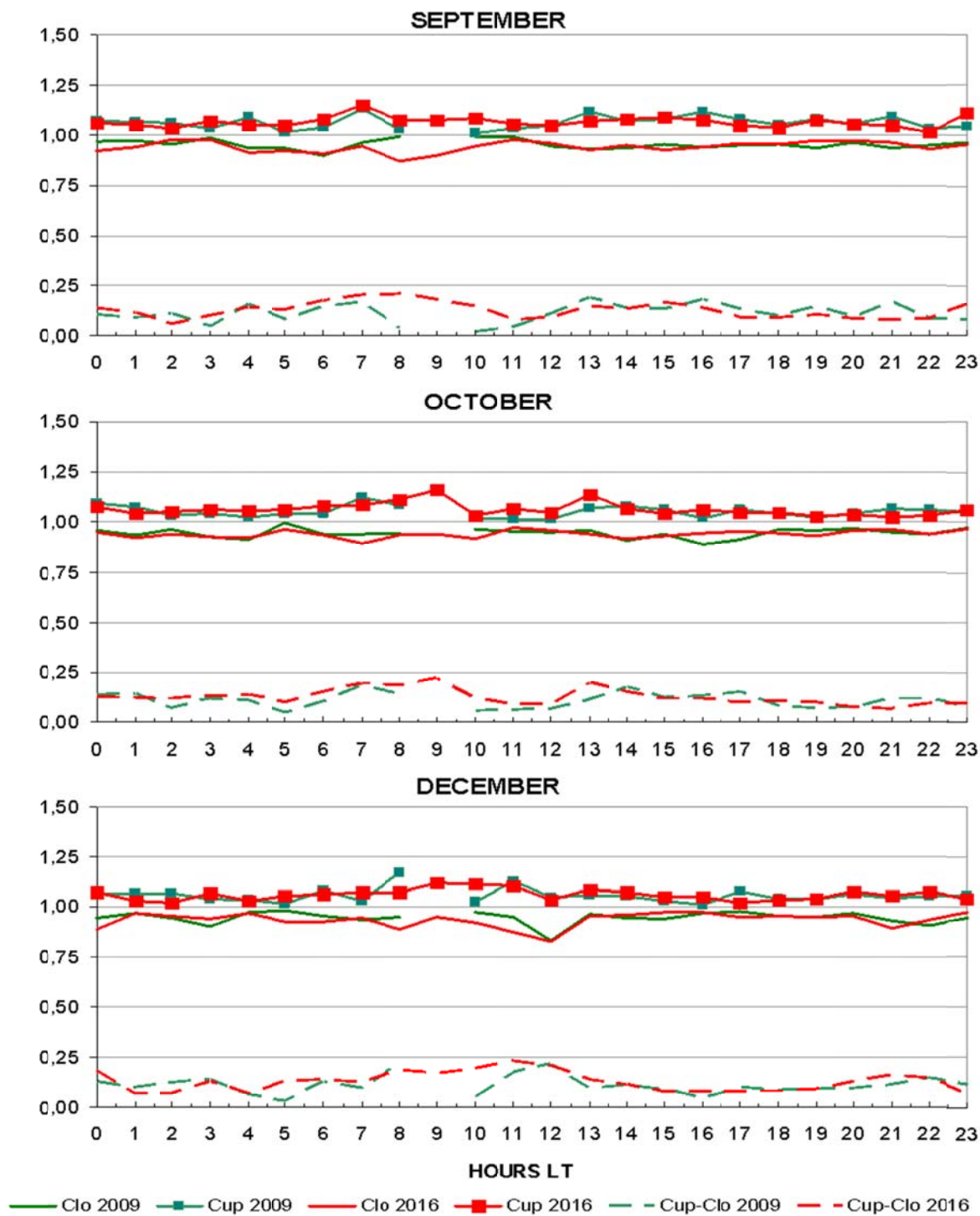

**Figure 13.** Temporal variability of upper and lower quartiles and interquartile range for hmF2 for one year of high solar activity (2016) and one of low solar activity (2009).

Whereas the Cr index for hmF2 shows that, in general, the variability during HSA and LSA is very similar; however, between 8 and 10 LT Cr for HSA is higher and a peak is observed during autumn and winter. In July and August, a second peak is present at 17 – 18 LT, while for spring months it is possible to see less intense peaks, at 8 LT on September, at 9 and at 13 LT on October and at 11 LT on December. During LSA, small peaks are observed, in March at 5, 7 and 17 LT, in April at 2 and 8 LT, in June at 17 LT, in July at 7 and 17 LT, in September at 7, 13, 16 and 21 LT, in October at 7 and 14 LT and in December at 12 LT. In contrast to foF2, the variability of hmF2 is not bigger during the night, and there is no significant difference between the three seasons analyzed. The Cr index for foF2 is higher than that for hmF2, the former has values of around 0.5 during the night especially in equinox LSA while the latter is generally smaller than 0.25 with only a few exceptions in April, May and July where there is a peak of ~0.3 during HSA.

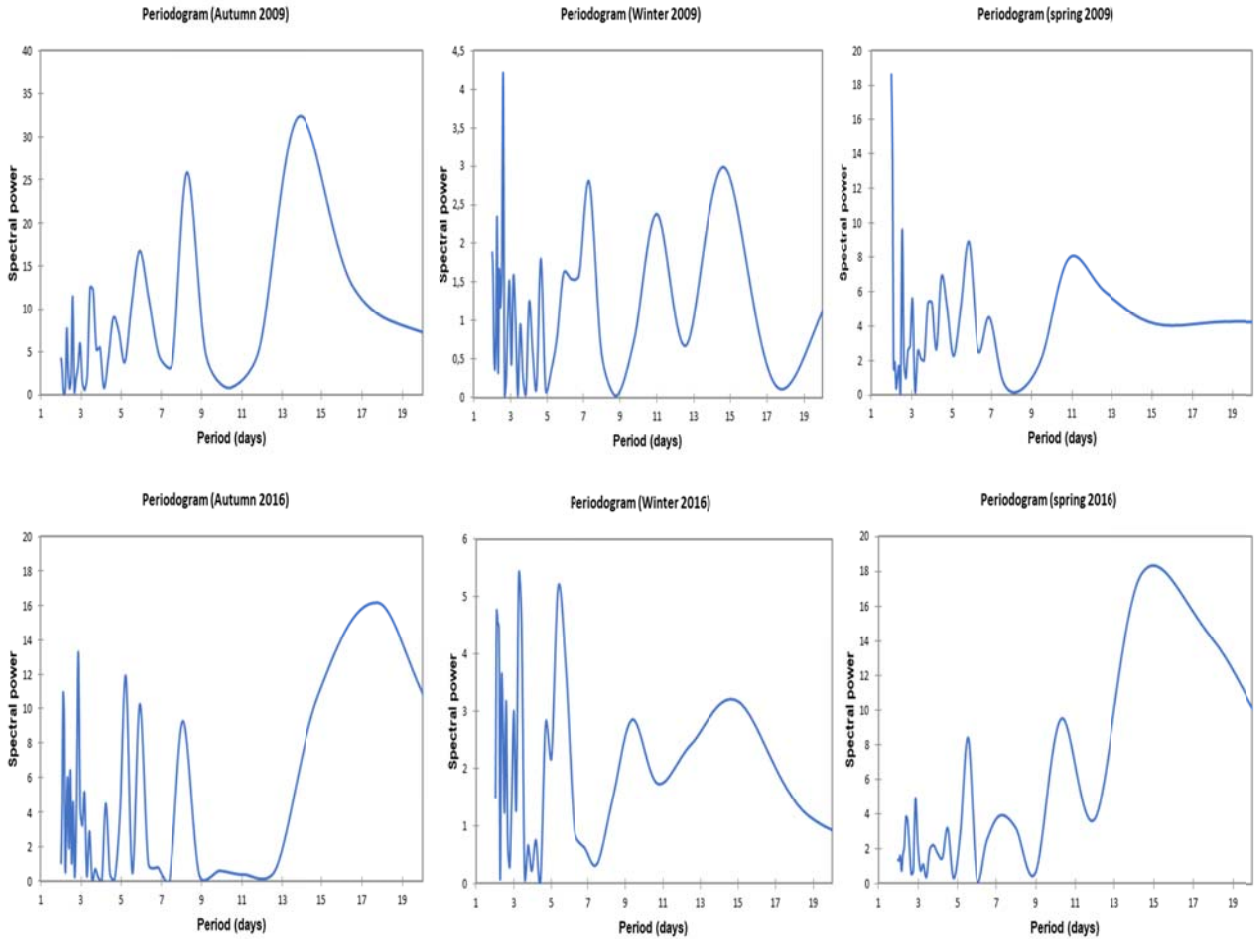

**Figure 14.** Lomb-Scargle periodogram for autumn, winter and spring 2009 (low solar activity) and 2016 (high solar activity).

In addition, an analysis of the ionospheric behavior during aphelion and perihelion is performed. Small differences were expected since the magnitude of the change in the Earth-Sun distance between these two positions is approximately 3%. A comparison of the interquartile difference during the aphelion (July) and the perihelion (December is used because there is no data for January) suggests that the variability of foF2 is slightly higher during the aphelion for high and low solar activity, specially between 12 and 18 LT, while the difference in the variability of hmF2 is not significant.

### 3-2. Atmospheric waves
Additionally, foF2 data was used to identify the ionospheric effects of waves in the neutral atmosphere. Figure 14 shows the Lomb-Scargle periodogram for autumn, winter and spring for LSA (2009) and HSA (2016) for a period range of 1 – 20 days. It is possible to see, planetary wave-like oscillations during all the seasons studied. For LSA, in autumn waves periods are 13.8, 8.3, 6, 3.6 and 2.6 days; in winter 2.6, 14.7, 7.3, 11 and 4.6 days; and in spring 2, 6 and 11 days. For HSA, in autumn the periods observed are 18, 3, 5.2, 2 and 8 days; in winter 3.3, 5.4, 2, 15 and 9.5 days; and in spring 14.4, 10.3, 5.5 and 3 days. For both periods of solar activity and for the three seasons, significant planetary wave spectral peaks at about 2 and 5 days are observed.

The typical relative amplitudes (amplitude divided by foF2) of planetary wave type oscillations are about 7% for LSA and 6% for HSA, but extreme values reach ∼ 16% for LSA and ∼10% for HSA. The planetary wave type oscillation amplitudes in foF2 are minima (of the order of 0.1 MHz) during winter specially for LSA and could be neglected for practical purpose like ionospheric predictions.

The analysis of foF2 data for 16 months of 2009 and 2016 also reveal the effects of tidal waves in the F2 region ionosphere. The periodograms present pronounced peaks at 20 – 23 hours for LSA and at 22 – 24 hours for HSA, these modulations are probably driven by solar tides. The behavior is similar for all the months analyzed. Besides, the 14 – 15-day modulation in foF2 observed in winter and spring during HSA and in winter LSA could be associated to a lunar semidiurnal tide.

### 3-3. Comparison with IRI-2016
Comparing the IRI model with observed data is useful for its improvement, especially at low latitudes. For that reason, the values of foF2 measured with the ionosonde of Tucumán and those obtained with the IRI-2016 model for nine months of 2009 and 2016 are compared. In this study, URSI is used, the ABT-2009 option is adopted for the bottom side thickness B0, and the NeQuick model is selected for the topside profile. Figures 15 and 16 show the temporal variability of the observed critical frequency (Exp foF2) and the modeled frequency (IRI foF2) for March, April, May, June, July, August, September, October and December 2009 (Low solar activity) and 2016 (High solar activity). It is observed that the IRI foF2 values are higher for the high solar activity period than for the low solar activity period, a behavior that agrees with the observations. On the other hand, the seasonal trend is similar; IRI foF2 and Exp foF2 are higher during spring and autumn than winter (semiannual anomaly). The peak values of IRI foF2 appear before that of Exp foF2 in most months and the largest time difference is 4 hours. IRI overestimates foF2 values between 5LT and approximately 15LT, while underestimates them between 16LT and 0LT for all months of 2009 and for March, April, August, September and October 2016. Between 0LT and 05LT, the agreement between observations and IRI values is very good, that is higher discrepancy exists for daytime than for nighttime and morning.

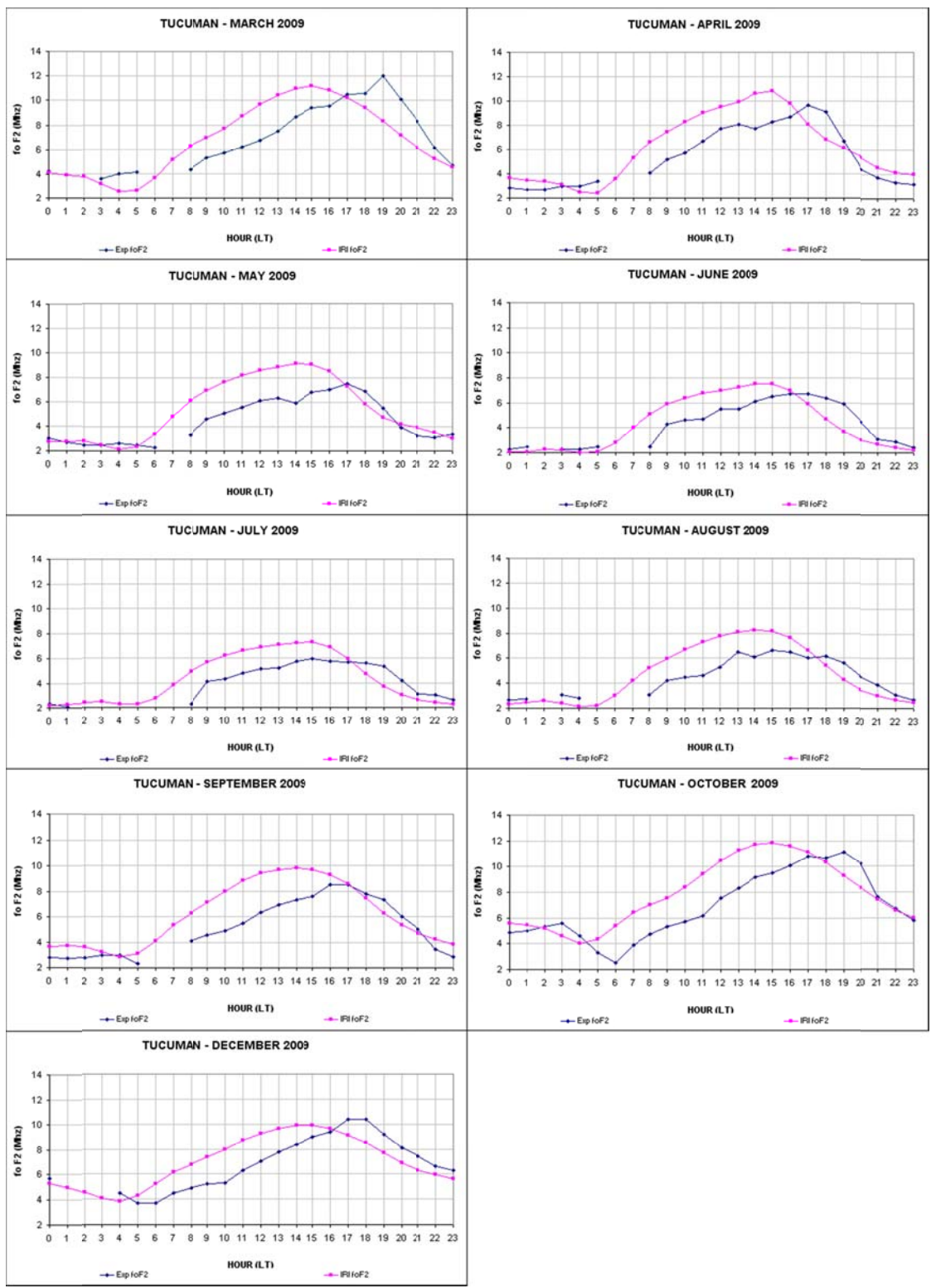

**Figure 15.** Time
series of foF2 measured in Tucuman (blue) and foF2 obtained with IRI model (pink) for nine months of a low solar
activity (LSA) year, 2009.

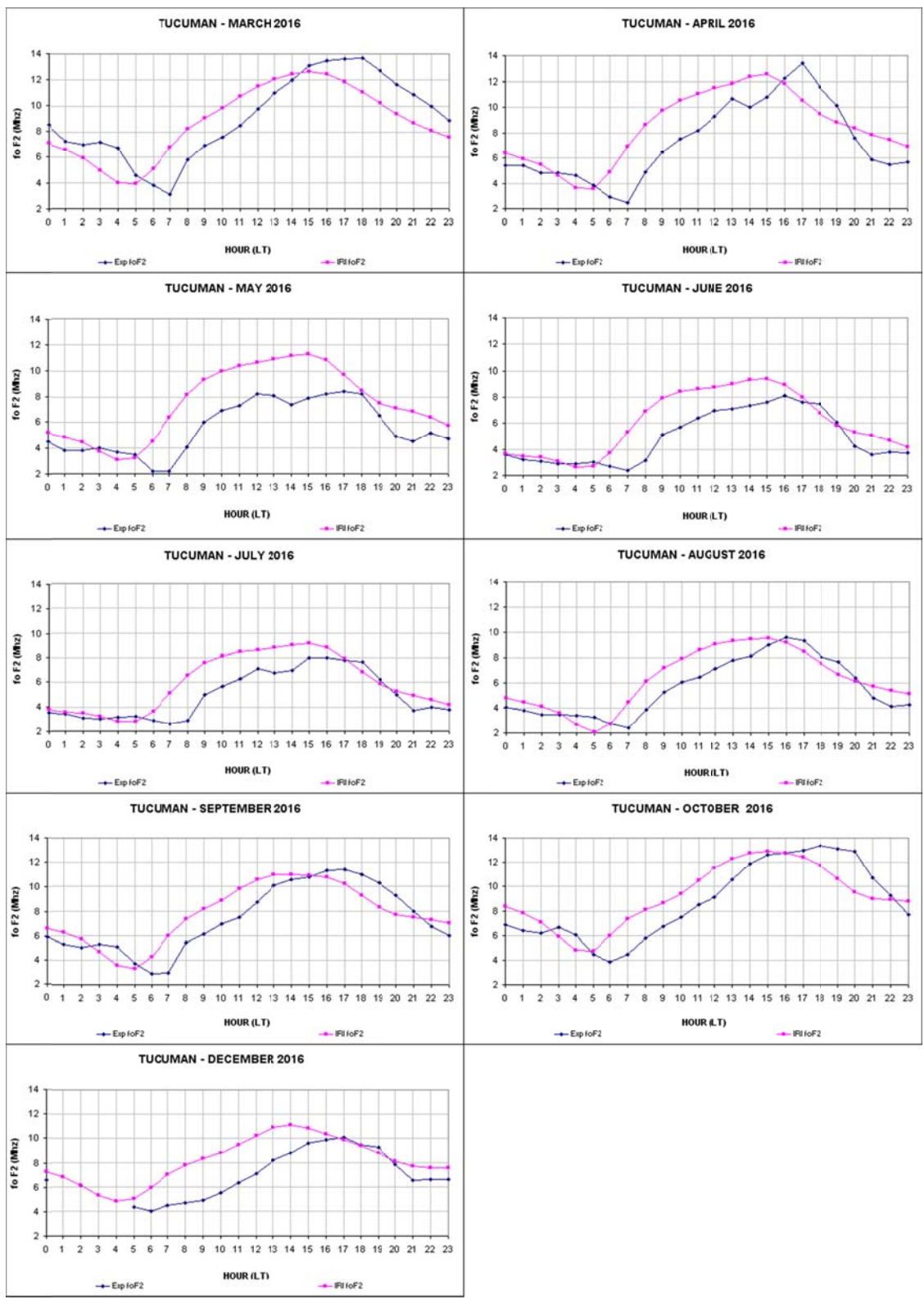

**Figure 16.** Time series of foF2 measured in Tucuman (blue) and foF2 obtained with IRI model (pink) for nine months of a high solar activity (HSA) year, 2016.

## 4. Conclusion

The present study examines the behavior of foF2 and hmF2 in three seasons during low and high solar activity. The data is obtained by an ionosonde measurements over a low latitude station at Tucuman-Argentina. (26.9 ° S, 294.6 ° E, magnetic latitude 15.5 ° S). Furthermore, the observed values are compared with the IRI-2016 modeled values using URSI option.

- A clear solar activity dependence is observed in foF2 and hmF2, both parameters increase with an increase in solar activity, except during December when foF2 data under low solar activity and high solar activity are very similar.

- The variability of foF2 is higher at low solar activity. This is not observed in hmF2, which present similar values under both periods.

- In general, the variability of foF2 is larger at night than during the day, this behavior is more pronounced during the high solar activity period. The variability of hmF2 is similar during all day, with a small increase between 8 and 10 LT under high solar activity.

- The semiannual anomaly is present irrespective of solar activity.

- In general, the variability of foF2 is higher than that of hmF2.

- Significant planetary wave spectral peaks at about 2 and 5 days are observed for high and low solar activity.

A possible reason of the larger variability under low solar activity is that electron density of the F2 layer is maximum during high solar activity, it is saturated and, therefore, the ionization is maximum, so the electron density is more stable than during low solar activity and the variability is smaller. The same happens during daytime, where due to the presence of the sun, the ionization is higher than during the nighttime, and there is less variability.

When comparing the observed values with the IRI-2016 model, it is observed that, IRI modeled values followed well the general trend for seasons under both high and low solar activity period, foF2 is higher in autumn and spring than in winter. The highest discrepancy occurs during daytime, where in general, IRI overestimates foF2. During post-sunset period, IRI underestimates foF2, and it shows a better agreement during nighttime. On the other hand, IRI modeled values are higher under high solar activity than during low solar activity, it coincides with the experimental observations.